%% file: samplepaper.tex
\documentclass[runningheads]{llncs}
\usepackage{graphicx}
\usepackage{tabularx}
\usepackage{dsfont}
\usepackage{amsmath}
\usepackage{amsthm}
\usepackage{graphicx}
\usepackage{tabularx}
\usepackage{hyperref}
\usepackage{listings}
\usepackage{subcaption}
\usepackage{multirow}
\usepackage{adjustbox}
\usepackage{lineno}
\usepackage{color}
\input{math_commands.tex}

\usepackage{footnote}
\makesavenoteenv{tabular}

\theoremstyle{definition}
\AtBeginDocument{%
  \providecommand\BibTeX{{%
    \normalfont B\kern-0.5em{\scshape i\kern-0.25em b}\kern-0.8em\TeX}}}

\begin{document}

\title{NeuralNDCG: Direct Optimisation of a Ranking Metric via Differentiable Relaxation of Sorting}
\titlerunning{NeuralNDCG: Direct Optimisation of Ranking Metrics...}
%
\author{Przemys\l{}aw Pobrotyn\thanks{Equal contribution.} \and
Rados\l{}aw Bia\l{}obrzeski\protect\footnotemark[1] }
\authorrunning{P. Pobrotyn and R. Bia\l{}obrzeski.}

\institute{ML Research at Allegro.pl \\
\email{mlr@allegro.pl} }

\maketitle              
\begin{abstract}
Learning to Rank (LTR) algorithms are usually evaluated using Information Retrieval metrics like Normalised Discounted Cumulative Gain (NDCG) or Mean Average Precision. As these metrics rely on sorting predicted items' scores (and thus, on items' ranks), their derivatives are either undefined or zero everywhere. This makes them unsuitable for gradient-based optimisation, which is the usual method of learning appropriate scoring functions. Commonly used LTR loss functions are only loosely related to the evaluation metrics, causing a mismatch between the optimisation objective and the evaluation criterion. In this paper, we address this mismatch by proposing NeuralNDCG, a novel differentiable approximation to NDCG. Since NDCG relies on the non-differentiable sorting operator, we obtain NeuralNDCG by relaxing that operator using NeuralSort, a differentiable approximation of sorting. As a result, we obtain a new ranking loss function which is an arbitrarily accurate approximation to the evaluation metric, thus closing the gap between the training and the evaluation of LTR models. We introduce two variants of the proposed loss function. Finally, the empirical evaluation shows that our proposed method outperforms previous work aimed at direct optimisation of NDCG and is competitive with the state\nobreakdash-of\nobreakdash-the\nobreakdash-art methods.

\keywords{Learning to Rank \and ranking metric optimisation \and NDCG approximation}
\end{abstract}
\section{Introduction}
Ranking is the problem of optimising, conditioned on some context, the ordering of a set of items in order to maximise a given metric. The metric is usually an Information Retrieval (IR) criterion chosen to correlate with user satisfaction. Learning to Rank (LTR) is a machine learning approach to ranking, concerned with learning the function which optimises the items' order from supervised data. In this work, without loss of generality, we assume our set of items are search results and the context in which we want to optimise their order is the user query.

Essentially, one would like to learn a function from search results into permutations. Since the space of all permutations grows factorially in the size of the search results set, the task of learning such a function directly becomes intractable. Thus, most common LTR algorithms resort to the approach known as \textit{score~\&~sort}. That is, instead of directly learning the correct permutation of the search results, one learns a \textit{scoring function} which predicts relevancies of individual items, in the form of real-valued scores. Items are then sorted in the descending order of the scores and thus produced ordering is evaluated using an IR metric of choice. Typically, scoring functions are implemented as either gradient boosted trees \cite{Friedman00greedyfunction} or Multilayer Perceptrons (MLP) \cite{rumelhart1985learning}. Recently, there has been work in using the Transformer \cite{attention} architecture as a scoring function \cite{context-aware-ranker}. In order to learn a good scoring function, one needs a tagged dataset of query-search results pairs together with ground truth relevancy of each search result (in the context of a given query), as well as a loss function. There has been extensive research into constructing appropriate loss functions for LTR (see \cite{Liu:2009:LRI:1618303.1618304} for an overview of the field). Such loss functions fall into one of three categories: \textit{pointwise}, \textit{pairwise} or \textit{listwise}.
Pointwise approaches treat the problem as a simple regression or classification of the ground truth relevancy for each individual search result, foregoing possible interactions between items. In pairwise approaches, pairs of items are considered as independent variables and the function is learned to correctly indicate the preference among the pair. Examples include RankNet \cite{RankNet}, LambdaRank \cite{LambdaRank} or LambdaMART \cite{burges2010ranknet}. However, IR metrics consider entire search results lists at once, unlike pointwise and pairwise algorithms. This mismatch has motivated listwise approaches, which compute the loss based on the scores of the entire list of search results. Two popular listwise approaches are ListNet \cite{ListNet}  and ListMLE~\cite{ListMLE}.



What these loss functions have in common is that they are either not connected or only loosely connected to the IR metrics used in the evaluation. The performance of LTR models is usually assessed using Normalised Discounted Cumulative Gain (NDCG) \cite{NDCG} or Mean Average Precision (MAP) \cite{MAP}. Since such metrics rely on sorting the ground truth labels according to the scores predicted by the scoring function, they are either not differentiable or flat everywhere and thus cannot be used for gradient-based optimisation of the scoring function. As a result, there is a mismatch between objectives optimised by the aforementioned pairwise or listwise losses and metrics used for the evaluation, even though it can be shown that some of such losses provide upper bounds of IR measures~\cite{SVM_map},~\cite{xu_adarank:boosting_2007}. On the other hand, as demonstrated in \cite{Qin2010}, under certain assumptions on the class of the scoring functions, direct optimisation of IR measures on a large training set is guaranteed to achieve high test performance on the same IR measure. Thus, attempts to bridge the gap between LTR optimisation objectives and discontinuous evaluation metrics are an important research direction.

In this work, we propose a novel approach to directly optimise NDCG by approximating the sorting operator with NeuralSort~\cite{NeuralSort}. Since the sorting operator is the source of discontinuity in NDCG (and other IR metrics), by substituting it with a differentiable approximation we obtain a smooth variant of the metric.
\newpage
The main contributions of the paper are:
\begin{itemize}
    \item We introduce NeuralNDCG, a novel smooth approximation of NDCG based on differentiable relaxation of the sorting operator. The variants of the proposed loss are discussed in Section \ref{Neural NDCG}.
    \item We evaluate a Context-Aware Ranker \cite {context-aware-ranker} trained with NeuralNDCG loss on Web30K~\cite{Web30K} and Istella \cite{istella} datasets. We demonstrate favourable performance of NeuralNDCG as compared to baselines. In particular, NeuralNDCG outperforms ApproxNDCG \cite{Qin2010}, a competing method for direct optimisation of NDCG.
    \item We provide an open-source Pytorch \cite{paszke2017automatic} implementation allowing for the reproduction of our results available as part of the open-source allRank framework\footnote{\href{https://github.com/allegro/allRank}{https://github.com/allegro/allRank}}.
\end{itemize}

The rest of the paper is organised as follows. In Section \ref{Related work}, we review the related literature. In Section \ref{Problem formulation} we formalise the problem of LTR. In Section \ref{Neural NDCG}, we recap NeuralSort and demonstrate how it can be used to construct a novel loss function, NeuralNDCG. In Section \ref{Experiments} we discuss our experimental setup and results. In the final Section \ref{Conlusions} we summarise our findings and discuss the possible future work.

\section{Related work}
\label{Related work}
As already mentioned in the introduction, most LTR approaches can be classified into one of three categories: pointwise, pairwise or listwise. For a comprehensive overview of the field and most common approaches, we refer the reader to \cite{Liu:2009:LRI:1618303.1618304}.

In this work, we are concerned with the direct optimisation of non-smooth IR measures. Methods for optimisation of such metrics can be broadly grouped into two categories. The methods in the first category try to optimise the upper bounds of IR metrics as surrogate loss functions. Examples include $\mathrm{SVM}^{\mathrm{map}}$ \cite{SVM_map} and $\mathrm{SVM}^{\mathrm{NDCG}}$ \cite{SVM_ndcg} which optimise upper bounds on $1 - \mathrm{MAP}$ and $1 - \mathrm{NDCG}$, respectively. On the other hand, ListNet was originally designed to minimise cross-entropy between predicted and ground truth top-one probability distributions, and as such its relation to NDCG was ill-understood. Only recently was it shown to bound NDCG and Mean Reciprocal Rank (MRR) for binary labels \cite{ListNetBinaryBound}. Further, a modification to ListNet was proposed in \cite{bruch2019alternative} for which it can be shown that it bounds NDCG also for the graded relevance labels. Popular methods like LambdaRank and LambdaMART forgo explicit formulation of the loss function and instead heuristically formulate the gradients based on NDCG considerations. Since the exact loss function is unknown, its theoretical relation to NDCG is difficult to analyse. 

The second category of methods aims to approximate an IR measure with a smooth function and directly optimise resulting surrogate function. Our method falls into this category. We propose to smooth-out NDCG by approximating non-continuous sorting operator used in the computation of that measure. Recent works proposing continuous approximation to sorting are already mentioned NeuralSort, SoDeep \cite{SoDeep} and smooth sorting as an Optimal Transport problem~\cite{OptimalTransportRanking}. We use NeuralSort for its firm mathematical foundation, the possibility to control the degree of approximation and ability to generalise beyond the maximum list length seen in training. SoDeep uses a deep neural network (DNN) and synthetic data to learn to approximate the sorting operator and as such lacks the aforementioned properties. Smooth sorting as Optimal Transport reports similar performance to NeuralSort at benchmark tasks and we aim to explore the use of it in NeuralNDCG in the future.
By replacing the sorting operator with its continuous approximation, we obtain NeuralNDCG, a differentiable approximation of the IR measure. Other notable methods for direct optimisation of NDCG include:
\begin{itemize}
    \item ApproxNDCG in which authors reformulated NDCG formula to involve summation over documents, not their ranks. As a result, they introduce a non-differentiable position function, which they approximate using a sigmoid. This loss has been recently revisited in a DNN setting in \cite{10.1145/3331184.3331347}.
    \item SoftRank \cite{SoftRank}, where authors propose to smooth scores returned by the scoring function with equal variance Gaussian distributions: thus deterministic scores become means of Gaussian score distributions. Subsequently, they derive an $O(n^3)$ algorithm to compute Rank-Binomial rank distributions using the smooth scores. Finally, NDCG is approximated by taking its expectation w.r.t. the rank distribution.
\end{itemize}

\section{Preliminaries}
\label{Problem formulation}
In this section, we formalise the problem and introduce the notation used throughout the paper. Let $(\vx, \vy) \in \mathcal{X}^n \times \mathbb{Z}_{\geq 0}^n$ be a training example consisting of a vector $\vx$ of $n$ items $x_i$, $1\leq i \leq n$, and a vector $\vy$ of corresponding non-negative integer relevance labels. Note that each item $x_i$ is itself a $d$-dimensional vector of numerical features, and should be thought of as representing a query-document pair. The set $\mathcal{X}$ is the space of all such vectors $\vx_i$. Thus, a pair $(\vx, \vy)$ represents a list of vectorised search results for a given query together with the corresponding ground truth relevancies. The dataset of all such pairs is denoted $\Psi$.  The goal of LTR is to find a scoring function $f: \mathcal{X}^n \rightarrow \mathbb{R}^n$ that maximises the chosen IR metric on $\Psi$. The scoring function is learned by minimising the empirical risk $\mathcal{L}(f) = \frac{1}{|\Psi|}\sum_{(\vx, \vy) \in \Psi}{\ell(\vy, \vs)}$ where $\ell(\cdot)$ is a loss function and $\vs = f(\vx)$ is the vector of predicted scores. As discussed earlier, in most LTR approaches there is a~mismatch between the loss function $\ell$ and the evaluation metric, causing a discrepancy between the learning procedure and its assessment. In this work, we focus on NDCG as our metric of choice and propose a new loss called NeuralNDCG, which bridges the gap between the training and the evaluation. Before we introduce NeuralNDCG, recall the definition of NDCG.

\begin{definition}
Let $(\vx, \vy) \in \mathcal{X}^n \times \mathbb{Z}_{\geq0}^n$ be a training example and assume the documents in $\vx$ have been ranked in the descending order of the scores computed using some scoring function $f$. Let $r_j$ denote the relevance of the document ranked at $j$-th position, $g(\cdot)$ denote a gain function and $d(\cdot)$ denote a discount function. Then, the Discounted Cumulative Gain at $k$-th position ($k \leq n$) is defined as
\begin{equation}
\mathrm{DCG}@k = \sum_{j=1}^{k} g(r_j) d(j)    
\end{equation}
and Normalised Discounted Cumulative Gain at $k$ is defined as
\begin{equation}
\mathrm{NDCG}@k = \frac{1}{\mathrm{maxDCG}@k}\mathrm{DCG}@k    
\end{equation}
where $\mathrm{maxDCG}@k$ is the maximum possible value of $\mathrm{DCG}@k$, computed by ordering the documents in $\vx$ by their decreasing ground truth relevancy.
\end{definition}

Note that, typically, the discount function $d(\cdot)$ and the gain function $g(\cdot)$ are given by $d(j) = \frac{1}{\log_2(j+1)}$ and $g(r_j) = 2^{r_j} - 1$, respectively.



\section{Loss formulation}
\label{Neural NDCG}
In this section we define NeuralNDCG, a novel differentiable approximation to NDCG. It relies on NeuralSort, a smooth relaxation of the sorting operator. We begin by recalling NeuralSort, proceed to define NeuralNDCG and discuss its possible variants.
\subsection{Sorting relaxation}
\label{sorting_relaxation}
Recall that sorting a list of scores $\vs$ is equivalent to left-multiplying a column vector of scores by the permutation matrix $P_{\texttt{sort}(\vs)}$ induced by permutation \texttt{sort($\vs$)} sorting the scores. Thus, in order to approximate the sorting operator, it is enough to approximate the induced permutation matrix. In \cite{NeuralSort}, the permutation matrix is approximated via a unimodal row stochastic matrix $\widehat{P}_{\texttt{sort}(\vs)}(\tau)$ given by:
\begin{equation} \label{eq:P_hat}
\widehat{P}_{\texttt{sort}(\vs)}[i, :](\tau) = \mathrm{softmax}[((n + 1 -2i)\vs - A_{\vs} \mathds{1}) / \tau]    
\end{equation}
where $A_{\vs}$ is the matrix of absolute pairwise differences of elements of $\vs$ such that $A_{\vs} [i, j] = |s_i - s_j|$, $\mathds{1}$ denotes the column vector of all ones and $\tau > 0$ is a temperature parameter controlling the accuracy of approximation. For brevity, for the remainder of the paper we refer to $\widehat{P}_{\texttt{sort}(\vs)}(\tau)$ simply as $\widehat{P}$.

Note that the temperature parameter $\tau$ allows to control the trade-off between the accuracy of the approximation and the variance of the gradients. Generally speaking, the lower the temperature, the better the approximation at the cost of a larger variance in the gradients. In fact, it is not difficult to demonstrate that:
\begin{equation}\label{eq:limit}
\lim_{\tau\to 0} \widehat{P}_{\texttt{sort}(\vs)}(\tau) = P_{\texttt{sort}(\vs)}    
\end{equation}
(see \cite{NeuralSort} for proof). This fact will come in handy once we define NeuralNDCG. 

An approximation of a permutation matrix by Equation \ref{eq:P_hat} is a deterministic function of the predicted scores. Authors of NeuralSort proposed also a stochastic version, by deriving a reparametrised sampler for a Plackett-Luce family of distributions. Essentially, they propose to perturb scores $s$ with a vector $g$ of i.i.d. Gumbel perturbations with zero mean and a fixed scale $\beta$ to obtain perturbed scores $\tilde{s} = \beta\log s + g$. Perturbed scores are then used in place of deterministic scores in the formula for $\widehat{P}$.

We experimented with both deterministic and stochastic approximations to sorting and found them to yield similar results. Thus, for brevity, in this work we focus on the deterministic variant.

\begin{table*}[h]
\begin{center}
\caption{Approximate sorting with NeuralSort. Given ground truth $\vy=[4, 2, 1, 0, 4, 3]$ and predicted scores $\vs~=~[0.5, 0.2, 0.1, 0.01, 0.65, 0.3]$, $\vy$ is sorted by $\widehat{P}$ for different values of $\tau$. Exact sorting is shown in the first row.}
\label{neural_sort_example}
\begin{tabular}{l|llllll|l}
& \multicolumn{6}{c|}{\bf{Quasi-sorted ground truth}}
& \bf{Sum after sorting} \\
\hline
$\lim_{\tau\to 0}$ & 4 & 4 & 3 & 2 & 1 & 0 & 14  \\
$\tau = 0.01$ & 4 & 4 & 3 & 2 & 0.99992 & 0.00012339 & 14.00004339 \\
$\tau = 0.1$ & 3.9995 & 3.8909 & 2.8239 & 1.9730 & 0.9989 & 0.3136 & 13.9998 \\
$\tau = 1$ & 3.3893 & 2.9820 & 2.4965 & 2.0191 & 1.6097 & 1.2815 & 10.388 \\
\hline
\end{tabular}
\end{center}
\end{table*}

\subsection{NeuralNDCG}
If the ground truth permutation is known, one could minimise the cross-entropy loss between the ground truth permutation matrix and its approximation given by $\widehat{P}$, as done in the experiments section in \cite{NeuralSort}. However, for many applications, including ranking, the exact ground truth permutation is not known. Relevance labels of individual items produce many possible valid ground truth permutation matrices. Thus, instead of optimising the cross-entropy loss, we use NeuralSort to introduce NeuralNDCG, a novel loss function appropriate for LTR. 

\begin{figure}[h]
    \centering
    \includegraphics[scale=0.6]{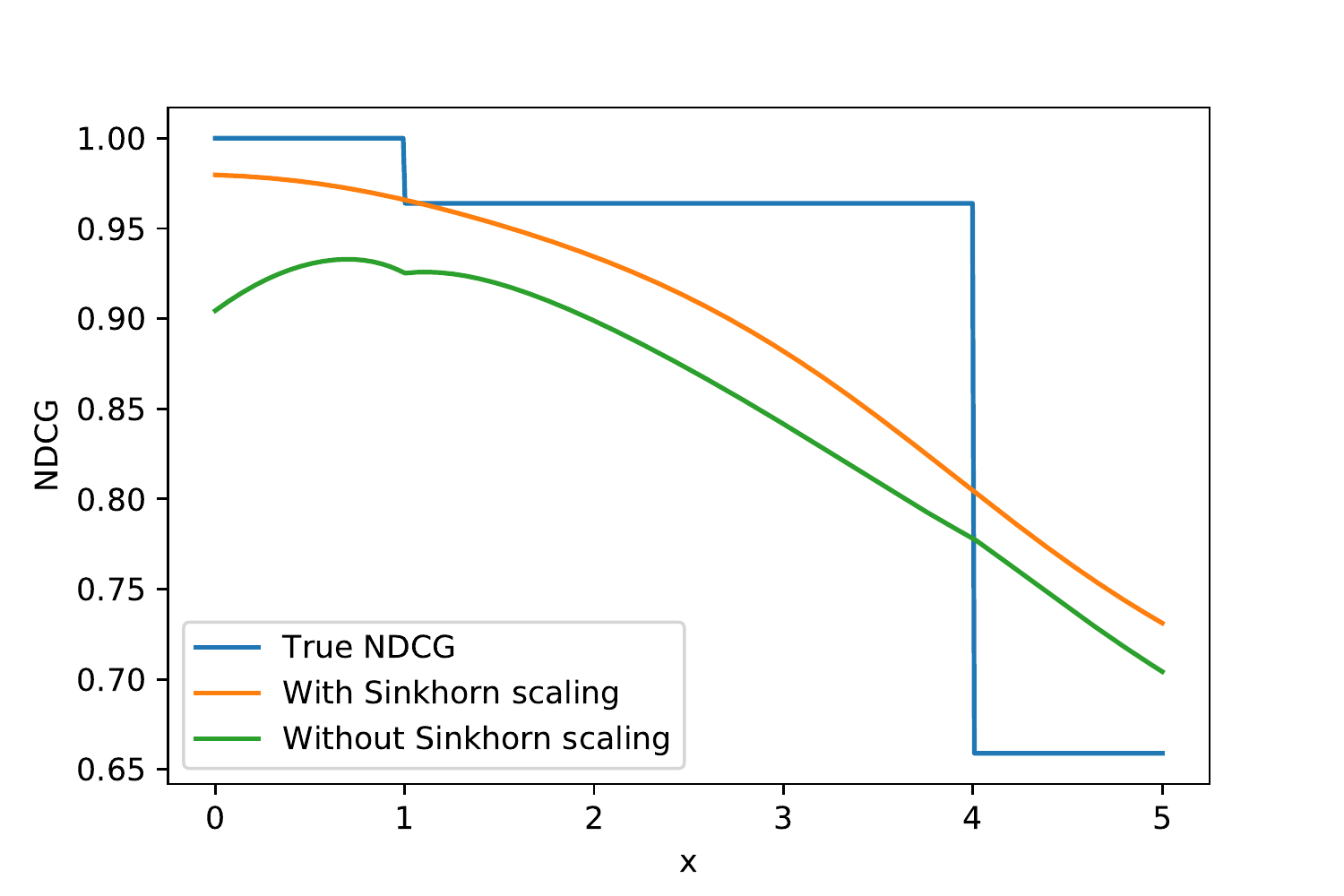}
    \caption{Given ground truth $y = [2, 1, 0, 0, 0]$ and a list of scores $\vs = [4, 1, 0, 0, x]$,  we vary the value of the score $x$ and plot resulting NDCG induced by the scores along with NeuralNDCG ($\tau = 1.0$) with and without Sinkhorn scaling of $\widehat{P}$.}
    \label{fig:approxNDCG_rescaling}
\end{figure}

Given a list of documents $\vx$, its corresponding vector of scores $\vs = f(\vx)$ and the ground truth labels $\vy$ we first find the approximate permutation matrix $\widehat{P}$ induced by the scores $\vs$ using Equation \ref{eq:P_hat}. We then apply the gain function $g$ to the vector of ground truths $\vy$ and obtain a vector $g(\vy)$ of gains per document.
We then left-multiply the column vector $g(\vy)$ of gains by $\widehat{P}$ and obtain an ``approximately" sorted version of the gains, $\widehat{g(\vy)}$. 
Another way to think of that approximate sorting is that the $k$-th row of $\widehat{P}$ gives weights of documents $x_i$ in the computation of gain at rank $k$ after sorting. Gain at rank $k$ is then the weighted sum of ground truth gains, weighted by the entries in the $k$-th row of $\widehat{P}$. Note that after the approximate sorting the actual integer values of ground truth gains become "distorted" and are not necessarily integers anymore (See Table \ref{neural_sort_example} for example). In particular, the sum of quasi-sorted gains $\widehat{g(\vy)}$ may differ from that of the original vector $g(\vy)$. This leads to a peculiar behaviour of NeuralNDCG near the discontinuities of true NDCG (Figure \ref{fig:approxNDCG_rescaling}), which may be potentially harmful for optimisation using Stochastic Gradient Descent \cite{Robbins&Monro:1951}. Since  $\widehat{P}$ is row-stochastic but not-necessarily column-stochastic (i.e. each column does not necessarily sum to one), an individual ground truth gain $g(\vy)_j$ may have corresponding weights in the rows of $\widehat{P}$ that do not sum to one (and, in particular, may sum to more than one), so it will overcontribute to the total sum of $\widehat{g(\vy)}$. To alleviate that problem, we additionally perform Sinkhorn scaling \cite{sinkhorn1964} on $\widehat{P}$ (i.e. we iteratively normalize all rows and columns until convergence\footnote{We stop the procedure after 30 iterations or when the maximum difference between row or column sum and one is less than $10^{-6}$, whatever happens first.})  before using it for quasi-sorting. This way, the columns also sum to one and the approximate sorting is smoothed-out (again, see Figure \ref{fig:approxNDCG_rescaling}). The remaining steps are identical to the computation of $\mathrm{NDCG}@k$, with the exception that the gain of the relevance function $r_j$ is replaced with the $j$-th coordinate of quasi-sorted gains $\widehat{g(\vy)}$. For the discount function $d$, we use the usual inverse logarithmic discount and for the gain function $g$ we used the usual power function. For the computation of the maxDCG, we use original ground truth labels $\vy$. To find NeuralNDCG at rank $k$, we simply truncate quasi-sorted gains to the $k$-th position and compute the maxDCG at $k$-th rank. 

We thus obtain the following formula for NeuralNDCG:

\begin{equation}
    \mathrm{NeuralNDCG}_k(\tau)(\vs, \vy) = N_k^{-1} \sum_{j=1}^{k} (\mathrm{scale}(\widehat{P})\cdot g(\vy))_j \cdot d(j)
\end{equation}
where $N_k^{-1}$ is the maxDCG at $k$-th rank, $\mathrm{scale}(\cdot)$ is Sinkhorn scaling and $g(\cdot)$ and $d(\cdot)$ are their gain and discount functions. Note that the summation is over the first $k$ ranks.

Finally, since the popular autograd libraries provide means to minimise a given loss functions, we use $(-1) \times \mathrm{NeuralNDCG}$ for optimisation.

\subsection{NeuralNDCG Transposed}
In the above formulation of NeuralNDCG, the summation is done over the ranks and gain at each rank is computed as a weighted sum of all gains, with weights given by the rows of $\widehat{P}$. We now provide an alternative formulation of NeuralNDCG, called NeuralNDCG Transposed ($\mathrm{NeuralNDCG}^T$ for short), where the summation is done over the documents, not their ranks.

As previously, let $\vx$ be a list of documents with corresponding scores $\vs$ and ground truth relevancies $\vy$. We begin by finding the approximate permutation matrix $\widehat{P}$. Since we want to sum over the documents and not their ranks, we need to find the weighted average of discounts per document, not the weighted average of gains per rank as before. To this end, we transpose $\widehat{P}$ to obtain an approximation $\widehat{P}^T$ of the inverse of the permutation matrix corresponding to sorting the documents $\vx$ by their corresponding scores $\vy$. Thus, $\widehat{P}^T$ can be thought of as an approximate \textit{unsorting} matrix - when applied to sorted documents (ranks), it will (approximately) recover their original ordering. Since $\widehat{P}$ is row-stochastic, $\widehat{P}^T$ is~column-stochastic. As we want to apply it by left-multiplication, we want it to be row-stochastic. Thus, similarly to before, we perform Sinkhorn scaling of $\widehat{P}^T$. After Sinkhorn scaling, the $k$-th row of $\widehat{P}^T$ can be thought of as giving the weights of different ranks when computing the discount of the $k$-th document. We can now find the vector of the weighted averages of discounts per document by computing $\widehat{P}^T \vd$, where $\vd$ is the vector of logarithmic discounts per rank ($\vd_j = d(j)$). Note that since we want to perform summation over the documents, not ranks, it is not enough to sum the first $k$ elements to truncate NDCG to the $k$-th position. Instead, the entries of the discounts vector $\vd$ corresponding to ranks $j > k$ are set to 0. This way, the documents which would end up at ranks $j > k$ after sorting end up having weighted discounts being close to 0, and equal to 0 in the limit of the temperature $\tau$. Thus, even though the summation is done over all documents, we still recover $\mathrm{NDCG}@k$.

Hence, $\mathrm{NeuralNDCG}^T$ is given by the following formula:

\begin{equation}
        {\mathrm{NeuralNDCG}^T}_k(\tau)(\vs, \vy) = N_k^{-1} \sum_{i=1}^{n} g(\vy_i) \cdot (\mathrm{scale}(\widehat{P}^T)\cdot \vd)_i
\end{equation}
where $N_k^{-1}$ is the maxDCG at $k$-th rank, $\mathrm{scale}(\cdot)$ is Sinkhorn scaling, $g(\cdot)$ is the gain function, $\vd$ is the vector of logarithmic discounts per rank set to 0 for ranks $j >k$, and the summation is done over all $n$ documents.


\subsection{Properties of NeuralNDCG}
\begin{figure}[h]
    \centering
    \includegraphics[scale=0.6]{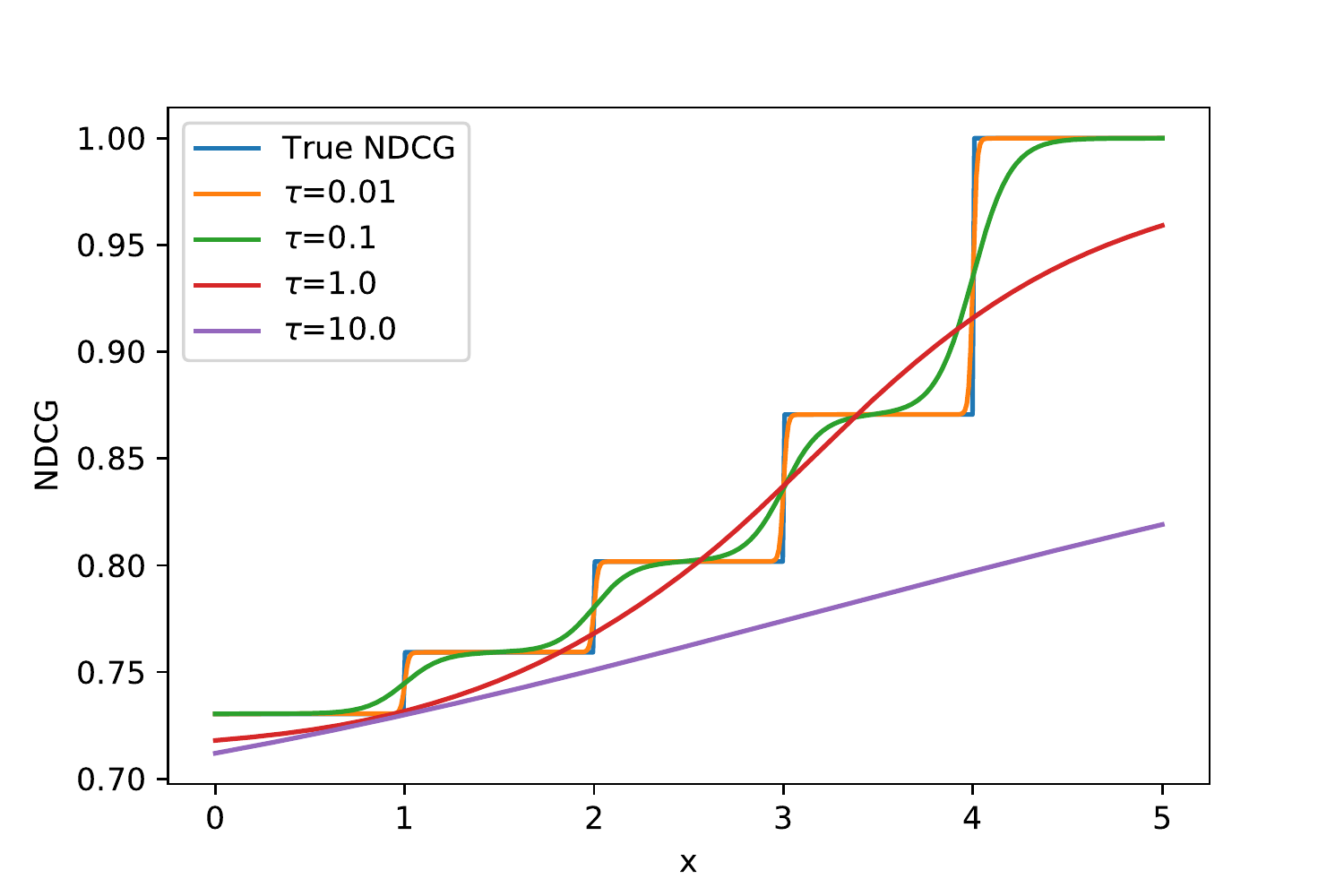}
    \caption{Given ground truth $\vy = [1,2,3,4,5]$ and a list of scores $\vs = [1, 2, 3, 4, x]$,  we vary the value of the score $x$ and plot resulting NDCG induced by the scores along with Deterministic NeuralNDCG for different temperatures $\tau$.}
    \label{fig:approxNDCG_temperature}
\end{figure}

By Equation \ref{eq:limit}, in the limit of the temperature, the approximate permutation matrix $\widehat{P}$ becomes the true permutation matrix $P$. Thus, as the temperature approaches zero, NeuralNDCG approaches true NDCG in both its variants. See Figure \ref{fig:approxNDCG_temperature} for examples of the effect of the temperature on the accuracy of the approximation.

Comparing to ApproxNDCG, our proposed approximation to NDCG showcases more favourable properties. We can easily compute NDCG at any rank position $k$, whereas in ApproxNDCG, one needs to further approximate the truncation function using an approximation of the position function. This approximation of an approximation leads to a compounding of errors. We deal away with that problem by using a single approximation of the permutation matrix. Furthermore, the approximation of the position function in ApproxNDCG is done using a sigmoid function, which may lead to the vanishing gradient problem. 

SoftRank suffers from a high computational complexity of $O(n^3)$: in order to compute all the derivatives required by the algorithm, a recursive computation is necessary. Authors relieve that cost by approximating all but a few of the Rank-Binomial distributions used, but at a cost of the accuracy of their solution. On the other hand, computation of $\widehat{P}$ is of $O(n^2)$ complexity. 

\section{Experiments}
\label{Experiments}
This section describes the experimental setup used to empirically verify the proposed loss functions. 
\subsection{Datasets}
Experiments were conducted on two datasets: Web30K and Istella \footnote{There are a few variants of this dataset, we used the Istella full dataset.}. Both datasets consists of queries and associated search results. Each query-document pair is represented by a real-valued feature vector and has an associated graded relevance on the scale from 0 (irrelevant) to 4 (highly relevant). For both datasets, we standardise the features, log-transforming selected ones, before feeding them to the learning algorithm.  Since the lengths of search results lists in the datasets are  unequal,  we  padded  or  subsampled  to  equal  length  for  training, but used the full list length for evaluation.
Web30K comes split into five folds. However, following the common practice in the  field,  we  report  results  obtained  on  Fold  1  of  the  data.  We  used  60\%  of the  data  for  training,  20\%  for  validation  and  hyperparameter  tuning  and  the remaining 20\% for testing. Istella datasets comes partition into a training and a test fold according to a 80\%-20\% schema. We additionally split the training data into training and validation data to obtain a 60\%/20\%/20\% split, similarly to Web30K. We tune the hyperparameters of our models on the validation data and report performance on the test set, having trained the best models on the full training fold.
In both datasets there are a number of queries for which the associated search results list contains no relevant documents (i.e. all documents have label 0). We refer to these queries as \textit{empty queries.} For such queries, the NDCG of their list of results can be arbitrarily set to either 0 or 1. To allow for a fair comparison with the current state of the art, we followed LightGBM \cite{ligthbGBM} implementation of setting NDCG of such lists to 1 during the evaluation.
Table~\ref{datasets} summaries the characteristics of the datasets used.

\begin{table}[]
\begin{center}
\caption{Dataset statistics}
\label{datasets}
\begin{tabular}{l|l|l|l|l}
Dataset & Features & Queries in training & Queries in test & Empty queries \\
\hline
Web30K  & 136      & 18919               & 6306            & 982           \\
Istella & 220      & 23219               & 9799            & 50       \\
\hline
\end{tabular}
\end{center}
\end{table}

\subsection{Scoring function}
For the scoring function $f$, we used the Context-Aware Ranker, a ranking model based on the Transformer architecture. The model can be thought of as the encoder part of the Transformer, taking raw features of items present in the same list as input and outputting a real-valued score for each item. Given the ubiquity of Transformer-based models in the literature, we refer to reader to \cite{context-aware-ranker} for the details of the architecture used. Compared to the original network described in \cite{context-aware-ranker}, we used smaller architectures. 
For both datasets, we used an architecture consisting of 2 encoder blocks of a single attention head each, with a hidden dimension of 384. The dimensionality of initial fully-connected layer was set to 96 for models trained on Web30K and 128 for models trained on Istella. We did not apply any activation on the output except for NeuralNDCG and NeuralNDCG$^T$. It exhibited suboptimal performance without any nonlinear output activation function and, in this case, we applied Tanh to the output. For both datasets, the same architectures were used across all loss functions.

\subsection{Training hyperparameters}
In all cases, we used Adam \cite{Adam} optimiser and set the learning rate to $0.001$. The batch size was set to 64 (Web30K) or 110 (Istella) and search results lists were truncated or padded to the length of 240 when training. We trained the networks for 100 epochs, decaying the learning rate by the factor of $0.1$ after 50 epochs.
\subsection{Loss functions}
We compared the performance of variants of NeuralNDCG against the following loss functions. For a pointwise baseline, we used a simple RMSE of predicted relevancy. Specifically, the output of the network $f$ is passed through a sigmoid function and then multiplied by the number of relevance levels. The root mean squared difference of this score and the ground truth relevance is the loss. Pairwise losses we compared with consist of RankNet and LambdaRank. Similarly to NeuralNDCG, these losses support training with a specific rank cutoff. We thus train models with these losses at ranks $5$, $10$ and at the maximum rank. The two most popular listwise losses are ListNet and ListMLE, and we, too, included them in our evaluation. Finally, the other method of direct optimisation of NDCG which we compared with was ApproxNDCG. We did not compare with SoftRank, as its $O(n^3)$ complexity proved prohibitive. We tuned ApproxNDCG and NeuralNDCG smoothness hyperparameters for optimal performance on the test set. Both ApproxNDCG's $\alpha$ and NeuralNDCG's $\tau$ parameter were set to 1 as other values in the $[0.01;100]$ interval did not show any improvement.

\begin{table}[h]
\caption{Test NDCG on Web30K and Istella. Boldface is the best performing loss column-wise.}
\label{web30k_test_fold1}
\begin{center}
\begin{tabular}{l|cc|cc}
\multicolumn{1}{c|}{\bf Loss}  
&\multicolumn{2}{c|}{\bf WEB30K} 
&\multicolumn{2}{c}{\bf Istella} \\
\hline
& NDCG@5  & NDCG@10  & NDCG@5  & NDCG@10  \\ 
\hline
NeuralNDCG@5 & 50.32 & 52.01 & 65.32 & 69.97  \\
NeuralNDCG@10 & 50.89 & 52.77 & 65.65  & 70.68  \\
NeuralNDCG@max & \textbf{51.56} & 53.46 & 65.69 & 70.55 \\
$\mathrm{NeuralNDCG}^T@5$ & 50.50 & 52.14 & 65.46 & 69.95  \\
$\mathrm{NeuralNDCG}^T@10$ & 50.85 & 52.70 & \textbf{66.02}  & 71.02  \\
$\mathrm{NeuralNDCG}^T$@\textrm{max} & 51.45 & \textbf{53.49} & 65.60 & 70.53 \\ 
ApproxNDCG & 49.07 & 50.90 & 63.14 & 67.94 \\
ListNet & 50.75 & 52.80 & 65.62 & 70.70 \\
ListMLE & 49.81 & 51.82 & 59.85 & 66.24 \\
RankNet@5 & 49.14 & 50.75 & 64.45 & 68.74 \\
RankNet@10 & 50.95 & 52.69 & 65.75 & 70.68  \\
RankNet@max & 49.84 & 51.82 & 64.57 & 70.37  \\
LambdaRank@5 & 48.70 & 50.10 & 63.50 & 67.75  \\
LambdaRank@10 & 49.66 & 51.34 & 65.21 & 69.82  \\
LambdaRank@max & 51.55 & 53.47 & 65.90 & \textbf{71.09}  \\
RMSE & 50.51 & 52.46 & 65.62 & 70.76 \\
\hline
\textrm{XGBoost} & 46.80 & 49.17 & 61.04 & 65.74 \\
\end{tabular}
\end{center}
\end{table}

\subsection{Results}
For both datasets, we report models performance in terms of NDCG@5 and NDCG@10. Results are collected in Table \ref{web30k_test_fold1}. Both NeuralNDCG variants in every rank cutoff setting outperform ApproxNDCG on both datasets in all metrics reported. Moreover, NeuralNDCG variants with specific rank cutoffs provide the best performance among all losses in both metrics on the WEB30K dataset and NDCG@5 on the Istella dataset. For reference, we also report the results of a GBDT model trained with XGBoost \cite{Chen:2016:XST:2939672.2939785} and objective rank:pairwise (as rank:ndcg is known to yield suboptimal results\footnote{For details, please visit \href{https://github.com/dmlc/xgboost/issues/6352}{https://github.com/dmlc/xgboost/issues/6352}.}).

\section{Conclusions}
\label{Conlusions}
In this work we introduced NeuralNDCG, a novel differentiable approximation of NDCG. By substituting the discontinuous sorting operator with NeuralSort, we obtain a robust, efficient and arbitrarily accurate approximation to NDCG. Not only does it enjoy favourable theoretical properties, but also proves to be effective in empirical evaluation, yielding competitive performance, on par with LambdaRank. This work can easily be extended to other rank-based metrics like MAP; a possibility we aim to explore in the future. Another interesting extension of this work would be the substitution of NeuralSort with another method of approximation of the sorting operator, most notably the method treating sorting as an Optimal Transport problem \cite{OptimalTransportRanking}.

\bibliographystyle{splncs04}
\bibliography{bibliography}


\end{document}

%% file: math_commands.tex
\usepackage{amsmath,amsfonts,bm}

\def\eqref#1{equation~\ref{#1}}

\def\1{\bm{1}}

\def\vd{{\bm{d}}}

\def\vs{{\bm{s}}}

\def\vx{{\bm{x}}}
\def\vy{{\bm{y}}}

\DeclareMathAlphabet{\mathsfit}{\encodingdefault}{\sfdefault}{m}{sl}
\SetMathAlphabet{\mathsfit}{bold}{\encodingdefault}{\sfdefault}{bx}{n}

